\documentclass[epj]{svjour}
\usepackage{graphics}
\input{psfig}
\newcommand{\vk}{\mbox {\boldmath $k$\unboldmath}}
\newcommand{\vq}{\mbox {\boldmath $q$\unboldmath}}
\def\nn{\nonumber}
\begin{document}
\authorrunning{Jun He {\it et al.}}
\titlerunning{Chiral constituent quark model study...}
\title{Chiral constituent quark model study of the process
$\gamma p \to \eta p$}
\author{Jun He\inst{1}, Bijan Saghai\inst{1},
Zhenping Li\inst{2}, Qiang Zhao\inst{3}, Johan Durand\inst{1}}
\institute{$^{1}$DAPNIA/SPhN, DSM, CEA/Saclay, 91191 Gif-sur-Yvette, France\\
 \ $^{2}$Department of Computer and Information Science, University of
Maryland, MD 20783, USA\\
\ $^{3}$Institute of High Energy Physics, Chinese Academy of Sciences,
Beijing, 100049, P.R. China}
\date{Received: date / Revised version: date}
\abstract{
A constituent quark model is developed for the reaction, allowing us
to investigate all available data for differential
cross sections as well as single polarization asymmetries (beam and target)
by including {\it all} of the PDG, one to four star, nucleon resonances
($S_{11}$, $P_{11}$, $P_{13}$, $D_{13}$, $D_{15}$, $F_{15}$, $F_{17}$,
$G_{17}$, $G_{19}$, $H_{19}$, $I_{1,11}$, and $K_{1,13}$).
Issues related to the missing resonances are also
briefly discussed by examining possible contributions from several
new resonances
($S_{11}$, $P_{11}$, $P_{13}$, $D_{13}$, $ D_{15}$, and $H_{1,11}$).
} 
\maketitle
\section{Introduction}
\label{intro}
Investigation of $\eta$-meson production {\it via} electromagnetic and
hadronic probes
offers access to fundamental information on hadron spectroscopy,
including the search for missing baryon resonances~\cite{Review}.
Extensive recent experimental efforts on $\eta$
photoproduction are opening a new era in this field.
The focus in this contribution is to study recent
$\gamma p \rightarrow \eta p$
data~\cite{Mainz,CLAS,ELSA,LNS,GRAAL,ELSA-pol,CB}
for $E_{\gamma}^{lab} \le$ 3 GeV
($W \equiv E_{total}^{cm} \le$ 2.6 GeV) within a chiral constituent
quark formalism~\cite{CQM1}, proven to be appropriate for the study of
meson photo- and electro-production~\cite{CQM2,CQM3,CQM4,CC06}, and
meson-nucleon scattering~\cite{IHEP} in the resonance region.

\section{Theoretical frame}
\label{sec:1}

The starting point of the meson photoproduction in the chiral quark
model is the low energy QCD Lagrangian~\cite{MG}
\begin{eqnarray}\label{eq:Lagrangian}
{\mathcal L}={\bar \psi} \left [ \gamma_{\mu} (i\partial^{\mu}+ V^\mu+\gamma_5
A^\mu)-m\right ] \psi + \dots,
\end{eqnarray}
where $\psi$ is the quark field in the $SU(3)$ symmetry,
$ V^\mu=(\xi^\dagger\partial_\mu\xi+\xi\partial_\mu\xi^\dagger)/2$
and
$A^\mu=i(\xi^\dagger \partial_{\mu} \xi -\xi\partial_{\mu} \xi^\dagger)/2$
are the vector and axial currents, respectively. The chiral transformation is
 $\xi=e^{i \phi_m / f_m}$, where $\phi_m$ is the Goldstone boson field and
$f_m$ the meson decay constant.
Then, the Lagrangian in Eq.~(\ref{eq:Lagrangian})
is invariant under the chiral transformation.
Therefore, there are four components for the photoproduction of
pseudoscalar mesons~\cite{CQM1} based on the QCD Lagrangian,
\begin{eqnarray}\label{eq:Mfi}
{\mathcal M}_{fi}&=&\langle N_f| H_{m,e}|N_i \rangle +
\sum_j\bigg \{ \frac {\langle N_f|H_m |N_j\rangle
\langle N_j |H_{e}|N_i\rangle }{E_i+\omega-E_j}+ \nn \\
&& \frac {\langle N_f|H_{e}|N_j\rangle \langle N_j|H_m
|N_i\rangle }{E_i-\omega_m-E_j}\bigg \}+{\mathcal M}_T,
\end{eqnarray}
where $|N_i\rangle$,
$|N_j\rangle$, and $|N_f\rangle$ stand for the initial,
intermediate, and final state baryons, respectively,
$\omega (\omega_{m})$ represents the energy of incoming (outgoing)
photons (mesons), and $H_m$ and $H_e$ are the pseudovector and electromagnetic
couplings at the tree level.

The first term in Eq.~(\ref{eq:Mfi}) is a seagull term.
The second and third terms correspond to the {\it s-} and {\it u-}channels,
respectively.
The last term is the {\it t-}channel contribution.

The contributions from  the {\it s-}channel resonances to the transition
matrix elements can be written as
\begin{eqnarray}\label{eq:MR}
{\mathcal M}_{N^*}=\frac {2M_{N^*}}{W^2-M_{N^*}(M_{N^*}-i\Gamma(q))}
e^{-\frac {{k}^2+{q}^2}{6\alpha^2_{ho}}}{\mathcal A}_{N^*},
\end{eqnarray}
with  $k=|\vk|$ ($q=|\vq|$) the momentum of the incoming photon
(outgoing meson), $W$ the total energy of
the system, $e^{- {({k}^2+{q}^2)}/{6\alpha^2_{ho}}}$ a form factor
in the harmonic oscillator basis with the parameter $\alpha^2_{ho}$
related to the harmonic oscillator strength in the wave-function,
and $M_{N^*}$ and $\Gamma(q)$ the mass and the total width of
the resonance, respectively.  The transition amplitudes ${\mathcal A}_{N^*}$
have been translated into the standard CGLN amplitudes in the harmonic
oscillator basis.

The contributions from each resonance
is determined by introducing~\cite{CQM2} a new set of
parameters $C_{{N^*}}$, and the substitution
\begin{eqnarray}\label{eq:C1}
{\mathcal A}_{N^*} \to C_{N^*} {\mathcal A}_{N^*},
\end{eqnarray}
so that,
\begin{eqnarray}\label{eq:C2}
{\mathcal M}_{N^*}^{exp} = C^2_{N^*}{\mathcal M}_{N^*}^{qm},
\end{eqnarray}
with ${\mathcal M}_{N^*}^{exp}$ the experimental value of
the observable, and ${\mathcal M}_{N^*}^{qm}$ calculated in the
quark model~\cite{CQM1}.
The $SU(6)\otimes O(3)$ symmetry predicts, e.g.
$C_{N^*}$~=~0.0 for the ${S_{11}(1650)} $, ${D_{13}(1700)}$, and
${D_{15}(1675)} $ resonances, and $C_{N^*}$~=~1.0 for other
$n \le$ 2 shell resonances.
Thus, the coefficients $C_{{N^*}}$ measure the discrepancies between
the theoretical results and the experimental data and show the extent
to which the $SU(6)\otimes O(3)$ symmetry is broken in the relevant process.
One of the main reasons that the $SU(6)\otimes O(3)$
symmetry is
broken is due to the configuration mixings caused by the one-gluon
exchange~\cite{IK}.
Here, the most relevant configuration mixings are those of the
two $S_{11}$ and the two $D_{13}$ states around 1.5 to 1.7 GeV. The
configuration mixings can be expressed in terms of the mixing angle
between the two $SU(6)\otimes O(3)$ states $|N(^2P_M)>$  and
$|N(^4P_M)>$, with the total quark spin 1/2 and 3/2.

In our previous investigations~\cite{CQM2}, the {\it s-}channel resonances with
masses above 2 GeV were treated as degenerate. In other words,
the transition amplitudes,
translated into the standard CGLN amplitudes were restricted to
harmonic oscillator shells $n \le 2$.
In the present work we extend that approach and derive explicitly the
amplitudes also for $n$=~3~to~6 shells~\cite{CQM5}. Moreover, we also include
{\it t-}channel contributions.

\section{Results and discussion}
\label{sec:2}
Using the formalism sketched above, we have investigated the cross-section and
single polarization observables for the process
$\gamma p \rightarrow \eta p$. In our model, non-resonant components include
nucleon pole term, {\it u-}channel contributions (treated as degenerate to the
harmonic oscillator shell $n$), and
{\it t-}channel contributions due to the $\rho$- and
$\omega$-exchanges~\cite{Nimai}.

The resonant part embodies the following $n$=1 to 6 shell nucleon resonances:
\begin{itemize}
 \item {\boldmath$ n$}{\bf =1:} $S_{11}(1535)$, $S_{11}(1650)$, $D_{13}(1520)$,
 $D_{13}(1700)$, and $D_{15}(1675)$;\smallskip
 \item {\boldmath$ n$}{\bf =2:} $P_{11}(1440)$, $P_{11}(1710)$, $P_{13}(1720)$,
 $P_{13}(1900)$,\\
$F_{15}(1680)$, $F_{15}(2000)$, and $F_{17}(1990)$;\smallskip
 \item {\boldmath$ n$}{\bf =3:} $\bf {S_{11}(1730)}$, $S_{11}(2090)$,
 $\bf {D_{13}(1850)}$, $D_{13}(2080)$, \\
$\bf {D_{15}(1950)}$, $D_{15}(2200)$, $G_{17}(2190)$,
 , and $G_{19}(2250)$;\smallskip
 \item {\boldmath$ n$}{\bf =4:} $\bf {P_{11}(1810)}$, $P_{11}(2100)$,
 $\bf {P_{13}(2170)}$,$H_{19}(2220)$, and $\bf {H_{1,11}(2200)}$;\smallskip
 \item {\boldmath$ n$}{\bf =5:} $I_{1,11}(2600)$;\smallskip
 \item {\boldmath$ n$}{\bf =6:} $K_{1,13}(2700)$.
\end{itemize}

Resonances considered here embody all of the 21 isospin-1/2 nucleon resonances
in the
PDG~\cite{PDG}, plus 6 new resonances reported in various publications.
Those resonances are given in the above list in bold character and refer
to the following works:
$\bf {S_{11}(1730)}$ \cite{Review,CQM2,CC06,S11,Gian},
$\bf {P_{11}(1810)}$ \cite{Gent,Anis,Sara},
$\bf {P_{13}(2170)}$ \cite{Review,CC06,Gian,BES},
$\bf {D_{13}(1850)}$ \cite{Review,CC06,Anis,Sara,D13},
$\bf {D_{15}(1950)}$ \cite{Anis,Sara},
$\bf {H_{1,11}(2200)}$ \cite{SAID}.

\noindent The masses attributed to those resonances are determined by the present work
and are compatible with other findings.
%

\begin{figure}
\psfig{figure=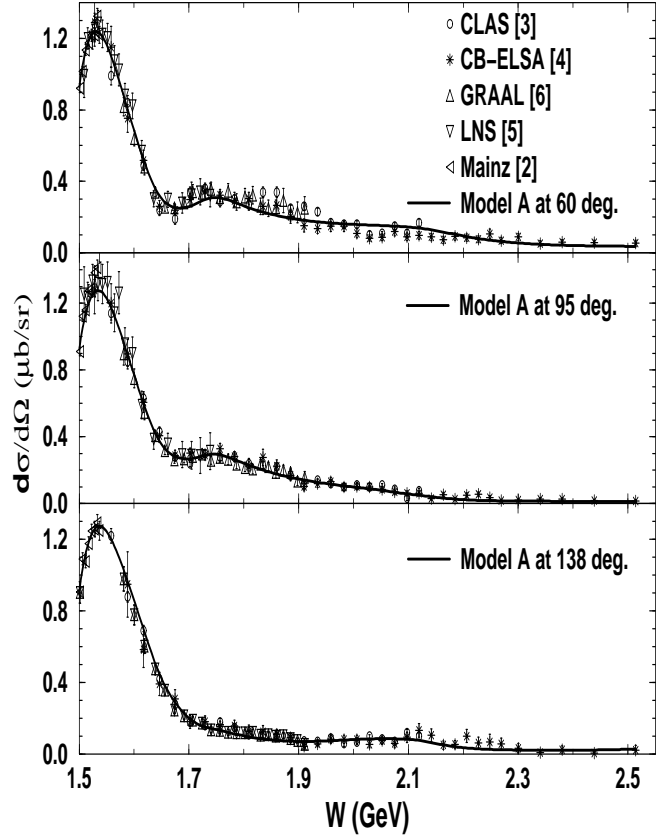,width=9.cm,height=12.cm
}
\caption{Excitation functions for the process $\gamma p \to \eta p$
at $\theta$ = 60$^\circ$, 95$^\circ$, and 138$^\circ$. Model $A$ is
explained in the text.}
\label{fig:ds}
\end{figure}

\begin{figure}
\psfig{figure=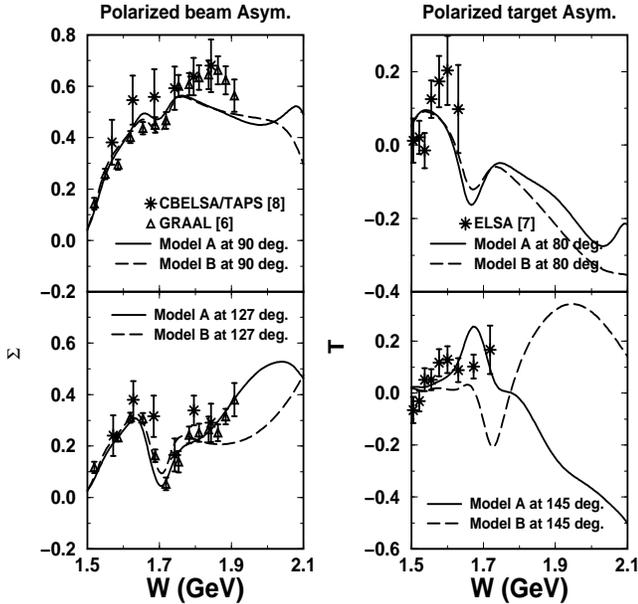,width=9.cm,height=8.cm
}
\caption{Polarization observables for the process $\gamma p \to \eta p$.
Left pannel: polarized beam asymmetries ($\Sigma$)
at $\theta$ = 90$^\circ$and 127$^\circ$.
Right pannel: polarized target asymmetries ($T$)
at $\theta$ = 80$^\circ$and 145$^\circ$. Models $A$ and $B$ are
explained in the text.}
\label{fig:pol}
\end{figure}

Because of lack of space, here we show excitation functions at three angles
(Fig.~\ref{fig:ds}),
polarized beam and target asymmetries at two angles (Fig.~\ref{fig:pol})
for each of the single polarization observables. (Consistencies among differential
cross-section data obtained at five facilities deserves to be underlined.)

Model $A$ has been obtained, using the CERN-MINUIT code, by fitting the data
with all $n$ = 1 to 6
shell resonances enumerated above. The adjustable parameters are the
$SU(6)\otimes O(3)$ symmetry breaking coefficients $C_{N^*}$
(Eq.~\ref{eq:C1}), and mass and width of the six new resonances.
The used data base contains 1588 differential
cross-sections \cite{Mainz,CLAS,ELSA,LNS,GRAAL},
184 polarized beam asymetries \cite{GRAAL,CB},
and 50 polarized target asymmetries \cite{ELSA-pol}.
The reduced $\chi ^2$ comes out to be 1.8.

The overall agreement between theory and experiment (Fig.~\ref{fig:ds})
is satisfactory. The extracted configuration mixing angles are
$\theta _S$ = -35$^\circ$ and $\theta _D$ = 15$^\circ$, close
enough to the Karl-Isgur model values~\cite{IK}.

Model $A$ reproduces reasonably well the single polarization observables
(Fig.~\ref{fig:pol}). Given the very limited number of data points for
the polarized target asymmetry ($T$), they do not put any significant
constraint on the adjustable parameters, hence, the curves can be considered
rather as predictions.

Starting from the full model $A$, we have switched off one resonance at a
time and checked the variation of the $\chi ^2$, without further minimizations.
In the model $A$, the dominant resonances come out to be the following nine
resonances:
$S_{11}(1535)$, $S_{11}(1650)$,
$\bf {S_{11}(1730)}$, $S_{11}(2090)$,
$P_{13}(1720)$, $P_{13}(1900)$,
$D_{13}(1520)$, $D_{13}(1700)$, and $F_{15}(1680)$. The highest $\chi ^2$
variations are observed in turning off the $S_{11}(1535)$, and to less extent,
the $D_{13}(1520)$. Among the six new resonances, only the $\bf {S_{11}(1730)}$
happens to play a significant role.

The model $B$ is obtained by fitting the data with merely those nine dominant
resonances plus the same non-resonant terms as in the model $A$. The reduced
$\chi ^2$ goes up from 1.8 (model $A$) to 2.1 (model $B$). For the differential
cross-sections depicted in Fig.~\ref{fig:ds}, the difference between the two models
$A$ and $B$ is not large enough to be visible in the figure. However, the single
polarization observables show significant sensitivities to the reaction
mechanism ingredients in models $A$ and $B$.The highest sensitivity is observed
in the case of polarized target asymmetry at backward angle and around
$W$ = 1.9 GeV.
Further data for this latter observable are highly desirable for a more
comprehensive understanding of the reaction mechanism.

In summary, we have developed a chiral quark model allowing us to include all
known and suggested nucleon resonances.
The data base, embodying 1822 data points, was fitted
successfully for differential cross-section and single polarization observables
(beam and target). Out of the 27 nucleon resonances included in the model, nine of
them play major roles in the reaction mechanism. One of those resonances
is a new one: $S_{11}$, for which we have extracted mass and width: $M$ = 1730
and $\Gamma$ = 100 MeV, compatible with other
findings~\cite{CQM2,CC06,S11,Gian}.

A more scrutiny investigation is in progress with respect to the sensitivity of
different data sets and/or energy ranges to the ingredients of our approach.
Afterwards, this elementary operator for the direct channel can be used
in a dynamical coupled-channel formalism~\cite{Bruno}, which embodies
$\pi N,~\pi \Delta,~\eta N,~\sigma N$, and $\rho N$ intermediate states.



\begin{thebibliography}{}
%
\bibitem{Review}
    S. Capstick and W. Roberts,
        {Prog. Part. Nucl. Phys.} \textbf{45}, (2000) 5241; and
         references therein.
%
\bibitem{Mainz}
        B. Krusche {et al.},
        {Phys. Rev. Lett.} \textbf{74}, (1995) 3736.
%
\bibitem{CLAS}
    M. Dugger {\it et al.} (CLAS Collaboration),
        Phys. Rev. Lett. \textbf{89}, (2002) 222002.
%
\bibitem{ELSA}
        V. Crede {\it et al.} (CB-ELSA Collaboration),
        {Phys. Rev. Lett.} \textbf{94}, (2005) 012004.
%
\bibitem{LNS}
        T. Nakabayashi {\it et al.},
        Phys. Rev. C \textbf{74}, (2006) 035202.
%
\bibitem{GRAAL}
    O. Bartalini {\it et al.} (GRAAL Collaboration),
    {Eur. Phys. J. A} \textbf{33}, (2007) 169.
%
\bibitem{ELSA-pol}
    A. Bock {\it et al.},
        {Phys. Rev. Lett.} \textbf{81}, (1998) 534;
%
\bibitem{CB}
         D. Elsner {\it et al.} (CB-ELSA and TAPS Collaborations),
    {Eur. Phys. J. A} \textbf{33}, (2007) 147.
%
\bibitem{CQM1}
    Z. Li, H. Ye, M. Lu,
        Phys. Rev. C \textbf{56}, (1997) {1099}.
%
\bibitem{CQM2}
        Z. Li and B. Saghai,
        {Nucl. Phys. A} \textbf{644}, (1998) {345};
%
         B. Saghai and Z. Li,
        {Eur. Phys. J. A} \textbf{11}, (2001) {217};
%
       B. Saghai and Z. Li,
       \textit{Proceedings of NSTAR 2002 Workshop on the Physics of Excited Nucleons},
       Pittsburgh, PA (USA), 2002; Editors S.A. Dytman and E.S. Swanson (World
       Scientific, New Jersey, 2003),
       arXiv: nucl-th/0305004.
%
\bibitem{CQM3}
         Q. Zhao, B. Saghai, Z. Li,
        {J. Phys. G} \textbf{28}, (2002) 1293.
%
\bibitem{CQM4}
         Q. Zhao, J.S. Al-Khalili, R.L. Workman,
        {Phys. Rev. C} \textbf{65}, (2002) 065204.
%
\bibitem{CC06}
        B. Juli\'a-D\'{\i}az, B. Saghai, T.-S.H. Lee, F. Tabakin,
        Phys. Rev. C \textbf{73}, (2006) 055204.
%
\bibitem{IHEP}
        Xian-Hui Zhong, Qiang Zhao, Jun He, Bijan Saghai,
        arXiv: nucl-th/0706.3543; nucl-th/0710.4212.
%
\bibitem{MG}
     A. Manohar and H. Georgi,
     Nucl. Phys. B \textbf{234}, (1984) 189.
%
\bibitem{IK}

        N. Isgur and G. Karl,
        {Phys. Lett. B} \textbf{72}, (1977) 109;
%
        N. Isgur, G. Karl, R. Koniuk,
        {Phys. Rev. Lett.} \textbf{41}, (1978) {1269};
%
        J. Chimza and G. Karl,
        Phys. Rev. D \textbf{68}, (2003) 054007.
%
\bibitem{CQM5}
        Jun He, Bijan Saghai, Zhenping Li, Qiang Zhao, Johan Durand,
        {\it manuscript in preparation}.
%
\bibitem{Nimai}
        M. Benmerrouche, N.C. Mukhopadhyay, J.F. Zhang,
        Phys. Rev. D \textbf{51}, (1995) 3237.
%
\bibitem{PDG}
          W. M. Yao {\it et al.},
          {J. Phys. G} {\bf 33}, (2006) 1.
%
\bibitem{S11}
        Z. Li and R. Workman,
        Phys. Rev. C \textbf{53}, (1996) R549;
%
       A. \v{S}varc and S. Ceci, arXiv: nucl-th/0009024;
%
       G.-Y Chen {\it et al.},
        Nucl. Phys. A \textbf{723}, (2003) 447;
%
         W.-T. Chiang{\it et al.},
         Phys. Rev. C {\bf 68}, (2003) 045202.
%
       V.~A.~Tryasuchev,
        Eur. Phys. J. A {\bf 22}, 97 (2004).
%
\bibitem{Gian}
        M.M. Giannini, E. Santopinto, and A. Vassallo,
        Eur. Phys. J. A {\bf 12}, (2001) 447.
%
\bibitem{Gent}
%
        T. Corthals, J. Ryckebusch, and T. Van Cauteren,
        Phys. Rev. C {\bf 73}, 045207 (2006).
%
\bibitem{Anis}
         A.V. Anisovich {\it et al.},
    {Eur. Phys. J. A} \textbf{25}, (2005) 427.
%
\bibitem{Sara}
         A.V. Sarantsev {\it et al.},
    {Eur. Phys. J. A} \textbf{25}, (2005) 441.
%
\bibitem{BES}
        M. Ablikim {\it et al.} (BES Collaboration),
         Phys. Rev. Lett. {\bf 97}, (2006) 202002.
%
        Sh. Fang (BES Collaboration),
        Int. J. Mod. Phys. A {\bf 21}, (2006) 839.
%
\bibitem{D13}
        T. Mart and C. Bennhold,
        Phys. Rev. C {\bf 61}, 012201 (2000);
%
        N.G. Kelkar, M. Nowakowski, K.P. Khemchandani, S.R. Jain,
           Nucl. Phys. A \textbf{730}, (2004) 121.
%
\bibitem{SAID}
     R.A. Arndt, W.J. Brisco, I.I. Strakovsky, and R.L. Workman,
     Phys. Rev. C {\bf 74}, 045205 (2006).
%
\bibitem{Bruno}
        B. Juli\'a-D\'{\i}az {\it et al.}, arXiv: nucl-th/0704.1615.
\end{thebibliography}
\end{document}